\documentclass[amsmath,amssymb,superscriptaddress,aps,prl,reprint,tightenlines]{revtex4-1}

\usepackage{graphicx}
\usepackage{float}
\usepackage{dcolumn}
\usepackage{bm}
\usepackage{xcolor}
\usepackage{natbib}
\usepackage{setspace}
\usepackage{amsmath}              
\usepackage{amssymb}              
\usepackage{amsfonts}             
\usepackage[amssymb]{SIunits}     
\usepackage[applemac]{inputenc}
\usepackage[english]{babel}
\newenvironment{rev}{\color{black}}{}

\begin{document}

\title{Minimal model of cellular symmetry breaking}
\author{Alexander Mietke}
\affiliation{Max Planck Institute for the Physics of Complex Systems, Dresden, Germany}
\affiliation{Chair of Scientific Computing for Systems Biology, Faculty of Computer Science, TU Dresden, Dresden, Germany}
\affiliation{Center for Systems Biology Dresden, Dresden, Germany}
\affiliation{Max Planck Institute of Molecular Cell Biology and Genetics, Dresden, Germany}
\author{V.~Jemseena}
\affiliation{International Centre for Theoretical Sciences, Tata Institute of Fundamental Research, Bengaluru, 
India}
\author{K.~Vijay~Kumar}
\affiliation{International Centre for Theoretical Sciences, Tata Institute of Fundamental Research, Bengaluru, 
India}
\author{Ivo F. Sbalzarini}
\affiliation{Chair of Scientific Computing for Systems Biology, Faculty of Computer Science, TU Dresden, Dresden, Germany}
\affiliation{Max Planck Institute of Molecular Cell Biology and Genetics, Dresden, Germany}
\affiliation{Center for Systems Biology Dresden, Dresden, Germany}
\affiliation{Cluster of Excellence Physics of Life, TU Dresden, Dresden, Germany}
\author{Frank J\"ulicher}
\affiliation{Max Planck Institute for the Physics of Complex Systems, Dresden, Germany}
\affiliation{Center for Systems Biology Dresden, Dresden, Germany}
\affiliation{Cluster of Excellence Physics of Life, TU Dresden, Dresden, Germany}
\affiliation{Corresponding author: julicher@pks.mpg.de}

\begin{abstract}
The cell cortex, a thin film of active material assembled below the cell membrane, plays a key role in cellular symmetry breaking processes such as cell polarity establishment and cell division. Here, we present a minimal model of the self-organization of the cell cortex that is based on a hydrodynamic theory of curved active surfaces. Active stresses on this surface are regulated by a diffusing molecular species. We show that coupling of the active surface to a passive bulk fluid  enables spontaneous  polarization and the formation of a contractile ring on the surface via mechano-chemical instabilities. We discuss the role of external fields in guiding such pattern formation. Our work reveals that key features of cellular symmetry breaking and cell division can emerge in a minimal model via general dynamic instabilities. 
\end{abstract}

\pacs{PACS}

\maketitle
The cortex of animal cells is a dynamically cross-linked polymer network located beneath the cell membrane~\cite{salb12}. It is involved in many important cellular symmetry-breaking events, such as the establishment of cell polarity~\cite{goeh11,call16} and cell division~\cite{whit83}. These processes typically involve cortical flows and cell shape changes, such that the cortex has to interact with material that surrounds it. Towards the inside of the cell, it is in contact with the cytoplasm, a crowded viscous fluid. By manipulating the cytoplasm mechanically, it has been shown that cytoplasmic flows can directly affect the dynamics of the cortex and the distribution of proteins within it~\cite{mitt18}. The reverse scenario, in which active cortical flows set the cytoplasmic fluid into motion, has also been observed~\cite{hird93}. This suggests that the cytoplasmic fluid is coupled to the dynamics of the cell cortex and \textit{vice versa}. 

The cell cortex has been successfully described as a thin active fluid film \cite{mayer10}. Many aspects of the cortex' emergent dynamics can be accounted for by considering its generic mechano-chemical organization~\cite{bois11}: The concentration of a diffusible chemical species regulates the amplitude of active stress, but also changes dynamically due to advection of the stress regulator by material flows. Spontaneous pattern formation in such self-organized active fluids has been studied on fixed domains with and without substrate friction~\cite{bois11,bane10,sark13,kuma14,moor14,sehr15,webe18} and on deforming surfaces in an environment with a homogeneous pressure~\cite{mietke18}. 

In this letter, we study a minimal model for the self-organization of an active surface that encloses a passive viscous fluid. A diffusing molecular species that regulates active tension on the surface provides a mechano-chemical feedback. We show that the coupling of the surface to the enclosed fluid gives rise to a hydrodynamic screening length 
that guides mechano-chemical instabilities to generate well-defined patterns on the surface. These patterns can govern shape changes and they can be oriented by external inhomogeneous signaling cues, which captures key features of symmetry-breaking events during important cellular processes.\\
 
We base our work on a simple hydrodynamic theory of a thin active fluid layer on a closed surface geometry~\cite{mietke18}. The surface $\Gamma$ is represented by a parametrization of surface position vectors $\mathbf{X}(s^1,s^2)$ by two generalized coordinates $s^1$, $s^2$. Tangent vectors
and unit surface normal are given by \hbox{$\mathbf{e}_i=\partial_i\mathbf{X}$}  ($\partial_i=\partial/\partial s^i$) and $\mathbf{n}=\mathbf{e}_1\times\mathbf{e}_2/|\mathbf{e}_1\times\mathbf{e}_2|$, respectively. 
Furthermore, we define the metric tensor $g_{ij}=\mathbf{e}_i\cdot\mathbf{e}_j$, the Levi-Civita tensor $\epsilon_{ij}={\bf n}\cdot(\mathbf{e}_i \times\mathbf{e}_j)$, and the curvature tensor $C_{ij}=-\mathbf{n}\cdot\partial_i\partial_j\mathbf{X}$.

The force and torque balance on the surface read~\cite{barth14}
\begin{align}
\nabla_i\mathbf{t}^i=-\mathbf{f}^{\text{ext}}\label{eq:FB}\\
\nabla_i\mathbf{m}^i=\mathbf{t}^i\times\mathbf{e}_i\label{eq:TB} \quad .
\end{align}
Here, we have introduced the surface stress $\mathbf{t}_i=t_{ij}\mathbf{e}^j+t_{n}^i\mathbf{n}$, the surface
moment $\mathbf{m}_i=m_{ij}\mathbf{e}^j+m_{n}^i\mathbf{n}$, and
$\nabla_i$ denotes the covariant derivative. The external force per unit area is denoted 
$\mathbf{f^{\text{ext}}}=f^{\text{ext}}_i\mathbf{e}^i+f^{\text{ext}}_n\mathbf{n}$. We do not include inertial forces and external torques. For simplicity, we do not consider deviatoric contributions to the moments. The tension and moment tensors in the surface can then be written as $t_{ij}=t_{ij}^e+t_{ij}^d$, $m_{ij}=m_{ij}^e$, $t^i_n=t^{i,e}_n+t^{i,d}_n$, and $m^i_n=m^{i,e}_n$, where the superscripts $e$ and $d$ refer to equilibrium and deviatoric contributions, respectively.

Equilibrium contributions can be obtained by considering a passive membrane with bending rigidity $\kappa$, spontaneous curvature $C_0$, and passive surface tension $\gamma$ as described by the Helfrich energy of a fluid membrane~\cite{helf73,mietke18}. In this case~\cite{salb17}:
\begin{align}
\ \hspace{-0.18cm}t_{ij}^e&=\gamma g_{ij}+\kappa\left(C^k_{\,k}-2C_0\right)\left(\left(C^k_{\,k}-2C_0\right)g_{ij}-2C_{ij}\right)\label{eq:HelfTens}\\
\ \hspace{-0.18cm}m_{ij}^e&=2\kappa\left(C^k_{\,k}-2C_0\right)\epsilon_{ij}\label{eq:HelfMom} \quad .
\end{align}  
For the deviatoric part of the tension tensor, we consider contributions from an isotropic active fluid film. In-plane material flows $\mathbf{v}_{\parallel}=v^i\mathbf{e}_i$ and surface deformations $\mathbf{v}_{\perp}=v_n\mathbf{n}$ contribute to the center-of-mass velocity $\mathbf{v}=\mathbf{v}_{\parallel}+\mathbf{v}_{\perp}$ of the active fluid film. The deviatoric tension tensor is given by~\cite{mietke18,salb17}
\begin{equation}\label{eq:ConstCort}
t^d_{ij}=2\eta_s\left(v_{ij}-\frac{1}{2}v^k_{\,k}g_{ij}\right)+\eta_bv^k_{\,k}\,g_{ij}+\xi_{ij} \quad .
\end{equation} 
Here,  $\eta_s$ and $\eta_b$ denote shear and bulk viscosity of the two-dimensional material, respectively and~$\xi_{ij}$ denotes an active tension. The strain rate tensor $v_{ij}=(\nabla_iv_j+\nabla_jv_i)/2+C_{ij}v_n$ captures the shear rate and area expansion of the thin material. 

With Eqs.~(\ref{eq:HelfTens}) and (\ref{eq:HelfMom}), the torque balance Eq.~(\ref{eq:TB}) implies $m_{n}^{i,e}=0$, $t^{i,d}_n=0$ and $t^{i,e}_n=\epsilon^i_{\;j}\nabla_k m^{kj,e}$, and we can express the force balance Eq.~(\ref{eq:FB}) as
\begin{align}
\nabla_it^{i,d}_{\ j}&=-f^{\text{ext}}_j\label{eq:FB_tang_specmain}\\
C^{ij}t_{ij}^d+f_n^e&=f^{\text{ext}}_n \label{eq:FB_norm_specmain} \quad.
\end{align} 
Here, we have defined $f_n^e=C^{ij}t^{e}_{ij}-\nabla_i t^{i,e}_n$ as the normal force exerted by a passive membrane~\cite{ouya89,suppPRL}. With the deviatoric tension tensor $t_{ij}^d$ from Eq.~(\ref{eq:ConstCort}), Eqs.~(\ref{eq:FB_tang_specmain}) and~(\ref{eq:FB_norm_specmain}) yield the hydrodynamic equations for the tangential and normal flow velocity, $\mathbf{v}_{\parallel}$ and $\mathbf{v}_{\perp}$, respectively.

The active surface encloses a passive bulk fluid. We describe the latter as an incompressible Stokes fluid ($\nabla\cdot\mathbf{u}=0$) obeying the force balance 
\begin{align}
\eta\Delta\mathbf{u}&=\nabla p,\label{eq:FB_CP}
\end{align}
where $\mathbf{u}$ denotes the passive flow field, $\eta$ is the shear viscosity, and $p$ denotes the hydrostatic pressure. To solve Eq.~(\ref{eq:FB_CP}), we impose no-slip and impermeability boundary conditions at the surface:
\begin{align}
\left.\mathbf{e}_i\cdot\mathbf{u}\right|_{\Gamma}&=v_i\label{eq:noslip}\\
\left.\mathbf{n}\cdot\mathbf{u}\right|_{\Gamma}&=v_n \quad.\label{eq:imperm}
\end{align}
The forces~$\mathbf{f^{\text{ext}}}$ in Eqs.~(\ref{eq:FB_tang_specmain}) and (\ref{eq:FB_norm_specmain}) result from viscous shear stresses that the passive fluid exerts on the surface. They are given by $\mathbf{f^{\text{ext}}}=-\mathbf{n}\left.\cdot\boldsymbol{\sigma}\right|_{\Gamma}$, where $\boldsymbol{\sigma}=\eta\left(\nabla\mathbf{u}+\nabla\mathbf{u}^T\right)-p\mathbb{I}$ is the stress tensor of the enclosed fluid. 

The equations for the active surface and the bulk fluid are combined with an advection-diffusion equation for stress regulator molecules of area concentration $c$ on the surface~\cite{salb17}:
\begin{equation}
\partial_t c+\nabla_i\left(c v^i\right)+C^k_{\,k}v_n c-D\Delta_{\Gamma}c =J_n\quad.\label{eq:ContEqSurf}
\end{equation}
Here, $\Delta_{\Gamma}=\nabla_i\nabla^i$ denotes the Laplace-Beltrami operator, $D$ is a diffusion coefficient and $J_n$ describes the exchange of molecules between the thin film and the enclosed fluid. It is given by
\begin{equation}
J_n=k_{\text{on}}\bar{c}|_{\Gamma}-k_{\text{off}}c \quad, \label{eq:Jn}
\end{equation}
where $k_{\text{on}}$ and $k_{\text{off}}$ denote rates for the recruitment of the stress regulator to and detachment from the surface, respectively. $\bar{c}$ is the volume concentration of molecules in the enclosed bulk fluid. For simplicity, we consider the case where the diffusion of the stress regulator in the enclosed fluid is fast compared to its exchange with the thin film. Then, the concentration~$\bar{c}$ is homogeneous with $d\bar{c}/dt = -V ^{-1}\oint_{\Gamma}dA J_n$, where $V$ is the volume of the enclosed bulk fluid.

Finally, the system is completed by a mechano-chemical feedback~\cite{mietke18}: The active tension~$\xi_{ij}$ in Eq.~(\ref{eq:ConstCort}) depends on the local surface concentration $c$ of the stress regulator molecules. We consider an active tension $\xi_{ij}=\xi f(c)g_{ij}$ that is isotropic within the surface, and the contractility $\xi$ is modulated by a function $f(c)$ with $\partial_cf(c)>0$ \cite{bois11}. Because of the mechanical coupling between the thin film and the enclosed fluid, given in Eqs.~(\ref{eq:noslip}) and (\ref{eq:imperm}), self-organized surface flows and deformations generated by active tension set the passive bulk fluid into motion.

Together, Eqs.~(\ref{eq:FB_tang_specmain}), (\ref{eq:FB_norm_specmain}), (\ref{eq:FB_CP}), and (\ref{eq:ContEqSurf}) represent a minimal model for cortical flows that are coupled to the cellular cytoplasm~\cite{mitt18}. This model has a simple stationary state in which the surface is given by a sphere of radius $R_0$, the surface concentration is homogeneous ($c=c_0$), and no flows exist ($\mathbf{v}=0,\mathbf{u}=0$). Two important time scales in this system are the time scale $\tau_c=\eta_b/\xi$ describing the advection-driven accumulation of stress regulator, and the diffusion time scale $\tau_D=R_0^2/D$. Then, $\text{Pe}=\tau_D/\tau_c=\xi R_0^2/(D\eta_b)$ can be identified as P\'eclet number characterizing the activity in the system~\cite{bois11,kuma14}. \\

\begin{figure}[!t]
	\centering	
\includegraphics[width=0.48\textwidth]{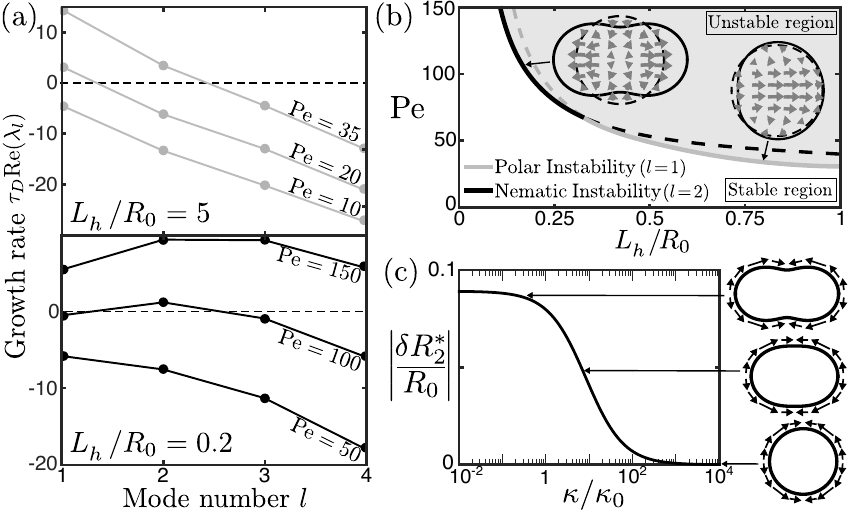} 
\caption{(a)~Eigenmode growth rates $\tau_D \lambda_l$ as a function of mode number $l$ for different P\'eclet numbers for larger (top) and smaller (bottom) \hbox{hydro}dynamic length $L_h=\eta_s/\eta$. Lines serve as guide to the eye. (b)~Linear stability diagram. The mode $l=1$ ($l=2$) becomes unstable first when moving across the gray (black) curve [Eq.~(\ref{eq:CritPeGen})]. Insets visualize unstable modes: Arrows depict bulk flows, outlines indicate perturbed shapes (large perturbation amplitude for visualization). The stability diagram is independent of bending rigidity~$\kappa$, spontaneous curvature~$C_0$, and surface tension~$\gamma$. (c)~Contributions of deformations $\delta R^*_{lm}$ to a critical eigenmode $\boldsymbol{\delta^*}_{lm}=(\delta R^*_{lm}/R_0,\delta c^*_{lm}/c_0)^T$ at $\text{Pe}=\text{Pe}^*_l$ as a function of $\kappa$~\cite{suppPRL}, shown here for $l=2, m=0$ ($\kappa_0=R_0^2\eta_b/\tau_D$). Parameters: $\kappa=0$ (a), $C_0=0$, $k_{\text{off}}\tau_D=10$, $\nu=1$, and $c_0\partial_cf(c_0)=1$.\label{fig:1}}
\end{figure}

We now discuss the linear stability of the homogeneous stationary state in which the active surface is given by a sphere. Using spherical harmonics $Y_{lm}(\theta,\varphi)$ ($l=0,1,...,\infty$; $m=-l,...,l$), where $\theta$ and $\varphi$ denote polar angle and azimuthal, respectively, as well as vector spherical harmonics $\boldsymbol{\Psi}^{(lm)}=R_0\nabla_{\Gamma}Y_{lm}$ and $\boldsymbol{\Phi}^{(lm)}=\hat{\mathbf{r}}\times\boldsymbol{\Psi}^{(lm)}$, we express shape, concentration, and flow perturbations as \smash{$\delta R=\sum_{l,m}\delta R_{lm}Y_{lm}$}, \smash{$\delta c=\sum_{l,m}\delta c_{lm}Y_{lm}$}, and $\delta\mathbf{v}_{\parallel}=\sum_{l,m}(\delta v^{(1)}_{lm}\boldsymbol{\Psi}^{(lm)}+\delta v^{(2)}_{lm}\boldsymbol{\Phi}^{(lm)})$~\cite{mietke18}.

We expand Eqs.~(\ref{eq:FB_tang_specmain}), (\ref{eq:FB_norm_specmain}), (\ref{eq:FB_CP}), and (\ref{eq:ContEqSurf}) to linear order in these fields~\cite{suppPRL,seyb18}. After eliminating the flow fields, the dynamics of each mode has the form $\frac{d}{dt}(\delta R_{lm},\delta c_{lm})^T={\cal J}_l(\delta R_{lm},\delta c_{lm})^T$, where ${\cal J}_l$ is the Jacobian. Its eigenvalues $\lambda_l$ are the growth rates of eigenmodes $\boldsymbol{\delta}_{lm}=(\delta R_{lm},\delta c_{lm})^T$. For vanishing or small P\'eclet number $\text{Pe}$, we have Re$(\lambda_l)<0$, and the steady state is stable (Fig.~\ref{fig:1}\,a). For increasing P\'eclet number and independently of the azimuthal mode number $m$, modes with $l\ge1$ become unstable at $\text{Pe}=\text{Pe}^*_l$, where
\begin{align} 
&\hspace{-0.15cm}\text{Pe}^*_l=\frac{1}{c_0\partial_cf(c_0)}\left(1+\frac{\tau_Dk_{\text{off}}}{l(l+1)}\right)\nonumber\\
&\hspace{0.45cm}\times\left[l(l+1)+\nu\left((l-1)(l+2)+(1+2l)\frac{R_0}{L_h}\right)\right]\label{eq:CritPeGen}
\end{align}
is the critical P\'eclet number for a mode $l$~\cite{suppPRL}. Here, we have defined the surface viscosity ratio $\nu=\eta_s/\eta_b$, as well as the hydrodynamic length $L_h=\eta_s/\eta$. Remarkably, $\text{Pe}^*_l$ is independent of bending rigidity $\kappa$,  spontaneous curvature~$C_0$, and surface tension~$\gamma$. Therefore, Eq.~(\ref{eq:CritPeGen}) equals the expression found in the limit of large $\kappa$, where the surface becomes a rigid sphere~\cite{suppPRL}. 

\begin{figure}[!t]
	\centering
\includegraphics[width=0.485\textwidth]{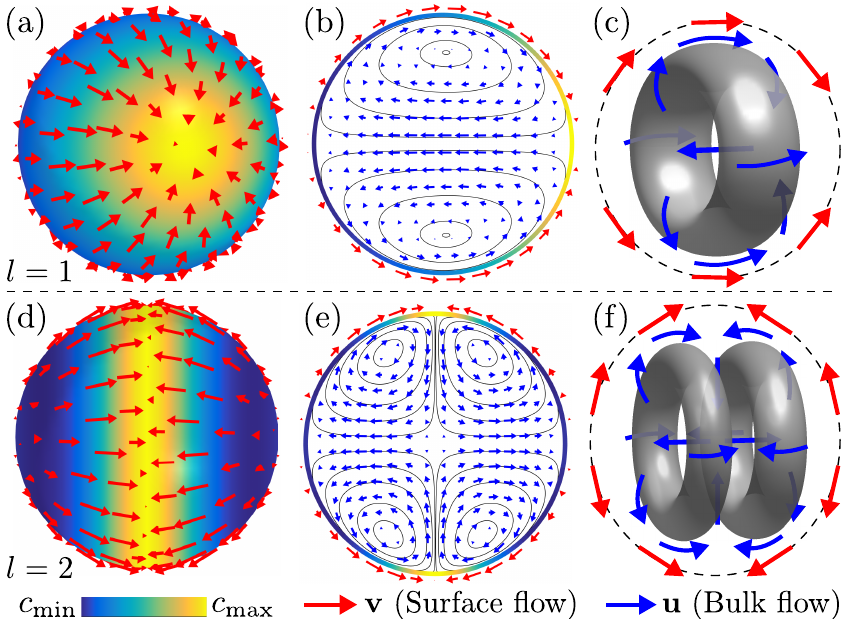} 
\caption{
(a)~An instability of the mode $l=1$ leads to a surface pattern with polar symmetry. (b)~Corresponding cross-sectional view parallel to the axis defined by the surface patterns. Black lines depict streamlines. (c)~Schematic representation of the global flow topology. Gray tori depict vortex rings, blue arrows indicate their direction of rotation. (d--f)~In regimes where $l=2$ is the only unstable mode, a contractile ring with nematic symmetry form. Parameters: $\text{Pe}=20$, $L_h/R_0=5$ ($l=1$); $\text{Pe}=100$, $L_h/R_0=0.2$ ($l=2$); $k\tau_D=10$, $\nu=1$, $\kappa\rightarrow\infty$ ($l=1,2$). Active tension is regulated by $f(c)=2c^2/(c_0^2+c^2)$, such that $c_0\partial_cf(c_0)=1$.\label{fig:2}}
\end{figure}

We now discuss Eq.~(\ref{eq:CritPeGen}) and key properties of the unstable modes in more detail. For small viscosities of the passive fluid, $\eta\lesssim\eta_s/R_0$, the mode $l=1$ becomes unstable first for increasing P\'eclet number. The instability of $l=1$ corresponds to a vectorial (polar) symmetry breaking. In the limit of large $L_h$ the viscosity of the surrounding  passive fluid can be neglected and we recover the result reported in~\cite{mietke18}. Interestingly, for finite turnover $k_{\text{off}}>0$ the nematic mode $l=2$ can become unstable, while $l=1$ is still stable (Fig.~\ref{fig:1}\,a, bottom). It follows from Eq.~(\ref{eq:CritPeGen}) that this can only occur for a small hydrodynamic length, $L_h\lesssim R_0$, corresponding to a regime where the stresses exerted by the enclosed passive fluid are significant. This implies that the hydrodynamic screening length $L_h$ plays a crucial role in selecting a specific wavelength for patterns on the surface. For finite surface tension $\gamma$ and bending rigidity~$\kappa$, the eigenmode associated with this instability is given by an ingression of the spherical surface along a ring of high stress regulator concentration. Critical eigenmodes~$\boldsymbol{\delta^*}_{lm}$ and a stability diagram of the spherical state as a function of P\'eclet number and hydrodynamic length are shown in~Fig.~\ref{fig:1}\,b. The homogeneous sphere is unstable in the gray shaded region. The polar ($l=1$) and nematic ($l=2$) instabilities, depicted by the gray and black curves, respectively, are given by Eq.~(\ref{eq:CritPeGen}). For $l\ge2$, critical eigenmodes depend on the bending rigidity~$\kappa$ and the surface tension~$\gamma$. In particular, contributions from deformations $\delta R_{lm}^*/R_0$ to the eigenmode vanish for large surface tension or bending rigidity~(Fig.~\ref{fig:1}\,c)~\cite{suppPRL}.
 
\ \\
To study the nonlinear dynamics beyond the discussed instabilities, we use numerical methods~\cite{suppPRL}. For simplicity, we consider the limit of large bending rigidity~$\kappa$, 
where the surface is not deformed. We first discuss the case $L_h/R_0=5$, where the polar mode becomes unstable first. Using a small random concentration perturbation as initial condition, the instability of $l=1$ leads to an axisymmetric steady-state pattern exhibiting a single patch of high stress regulator concentration (Fig.~\ref{fig:2}\,a--c). A cross section that contains the polar axis defined by this pattern reveals a hydrodynamic flow field with a backflow along the symmetry axis, driven by the active surface flows that maintain the pattern.
For \hbox{$L_h/R_0=0.2$} the mode $l=2$ can become unstable first for increasing P\'eclet number (Fig.~\ref{fig:1} b). In this case, a random perturbation leads to the formation
of a ring of high stress regulator concentration along the equator (Fig.~\ref{fig:2}\,d--f). This ring corresponds to a circumferential contractile ring of active tension that can constrict a deformable sphere. In this state, the passive fluid flow exhibits two toroidal vortex tubes, stacked orthogonally to the nematic axis and rotating in opposite directions. Further away from the instability threshold, numerical calculations reveal the existence of oscillatory 
 patterns in certain regimes~\cite{suppPRL}.
\\

\begin{figure}[!t]
	\centering
\includegraphics[width=0.48\textwidth]{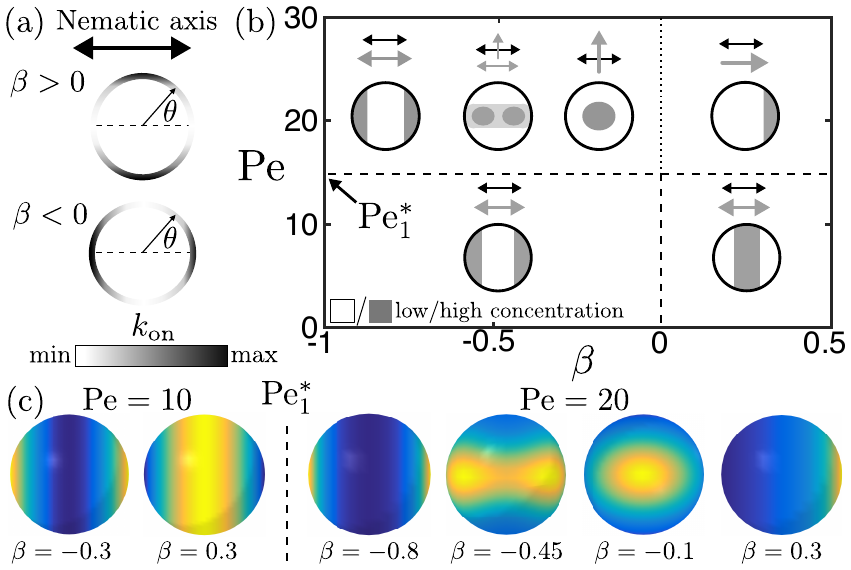} 
\caption{(a)~Cross-sectional views on recruitment rate~\smash{$k_{\text{on}}(\theta)$} [Eq.~(\ref{eq:kon})]. $\beta$ determines amplitude and sign of the nematic cue. (b)~Schematic representation of surface patterns and their orientation for varying P\'eclet number $\text{Pe}$ and strength of the nematic cue $\beta$ in a regime where $l=1$ becomes unstable first at $\text{Pe}=\text{Pe}^*_1$ if $\beta=0$ (dotted line). Gray arrows depict the orientation and symmetry axes defined by the surface patterns. For $\text{Pe}<\text{Pe}^*_1$ and $\beta\ne0$, steady-state surface patterns are dictated by~\smash{$k_{\text{on}}(\theta)$}. For $\text{Pe}>\text{Pe}^*_1$ and $\beta\ne0$, spontaneously forming patterns on the surface interact with the nematic cue. (c)~Representative steady-state concentration patterns obtained from numerical solutions. Qualitative color code as in Fig.~\ref{fig:2}\,a,d. Parameters: $L_h/R_0=5$, $k_{\text{off}}\tau_D=10$, and~$\nu=1$.\label{fig:ExtFieldSketch}} 
\end{figure}

The nematic instability provides a minimal model for the self-organized formation of a contractile ring that can drive constriction during cell division. In our model, the axis characterizing the contractile ring is defined by a spontaneous symmetry-breaking event. This is different from biological systems, where the contractile ring, and hence the division axis, are oriented along the mitotic spindle via biasing signaling cues~\cite{thery07,rodr15}. Also cell polarization, a process that is key to asymmetric cell divisions, depends on the coordination between spindle orientation and the biochemical organization of the cellular cortex~\cite{morin11}. In order to include such a symmetry-breaking bias in our model, we generalize the expression for the flux $J_n$ given in Eq.~(\ref{eq:Jn}) and consider an angle dependent recruitment rate $k_{\text{on}}(\theta)$ of stress regulator on the surface, described by 
\begin{align}
{k}_{\rm on}(\theta) &=k_{\rm on}^{(0)}\left[1+\beta\left(1-3\cos^2\theta\right)\right]\label{eq:kon}.
\end{align}  
The coefficient $\beta$ determines the strength and sign of the nematic bias. 
It varies in the interval \hbox{$[-1,1/2]$}, such that ${k}_{\rm on}(\theta)\ge0$. For $\beta>0$ ($\beta<0$), there is a recruitment of stress regulator predominantly to the equator region near $\theta=\pi/2$ (to the opposing poles at $\theta=0,\pi$) (Fig.~\ref{fig:ExtFieldSketch}\,a). 

We first consider the effects of the nematic cue on the self-organized pattern formation in the regime where the polar mode \hbox{$l=1$} becomes unstable first for increasing P\'eclet number (Fig~\ref{fig:ExtFieldSketch}). For $\text{Pe}<\text{Pe}^*_1$, the homogeneous state is stable in the absence of the cue ($\beta=0$), while $\beta\ne0$ leads to the formation of a concentration pattern with the nematic symmetry dictated by $k_{\text{on}}(\theta)$. For $\text{Pe}>\text{Pe}^*_1$, the polar instability in the presence of a nematic cue leads to more complex surface patterns that combine polar and nematic features (Fig.~\ref{fig:ExtFieldSketch}\,b). For $\beta>0$ a polar surface pattern forms, whose axis is oriented parallel to the axis of the nematic cue. For $\beta<0$, we can qualitatively distinguish three regimes. If $|\beta|$ is small, a single contractile patch forms, defining a polar axis that is oriented orthogonal to the nematic cue axis. If $|\beta|$ is increased, two local concentration maxima appear. If $|\beta|$ is increased further, the nematic cue dominates, leading to pattern with two patches of stress regulator at opposing poles aligned by the cue. Figure~\ref{fig:ExtFieldSketch}\,c shows examples of steady-state concentration patterns for these different cases. 

We also found that in the case of an instability with nematic symmetry (\hbox{$l=2$}), a cue with $\beta>0$ ensures that the nematic axis of the emerging contractile rings is reliably oriented parallel to the axis of the nematic cue. This captures the effect of a mitotic spindle with nematic symmetry orienting the contractile ring along the division axis.

\ \\
In this letter, we have studied the mechano-chemical self-organization of active fluid surfaces with spherical geometry. We have shown that the viscous forces exerted by a passive fluid on an enclosing active fluid film can control the formation of patterns with different symmetries. 

We have found that the active surface can undergo spontaneous symmetry-breaking instabilities toward patterns of concentration, flows, and deformations with polar or nematic symmetries, depending on the ratio $L_h/R_0$ of hydrodynamic screening length to sphere radius. For large ratios, polar patterns emerge, corresponding to the mode $l=1$, while patterns can be nematic, corresponding to $l=2$, for smaller ratios $L_h/R_0$. When decreasing the ratio $L_h/R_0$ even further, stationary patterns corresponding to higher harmonic modes $l>2$ can emerge. These have a polar symmetry for odd $l$ and a nematic symmetry if $l$ is even~\cite{suppPRL}.

For simplicity, we considered here an isotropic active tension. In general, an anisotropic contribution to the active tension of the form $\xi'_{ij}=\xi'f(c)C_{ij}$ exists. When taking such anisotropies in active tension into account, the results presented here do not change
qualitatively, but the critical P\'eclet number is altered for $l>1$~\cite{suppPRL}.

The emergence of mechano-chemical patterns presented in this work 
generalizes the one-dimensional contractile instabilities
described in~\cite{bois11} to curved surfaces. Instabilities on the surface of a sphere
discussed here capture key features of important cellular processes. The emergence of polar patterns resembles the establishment of cell polarity by active processes in the cell cortex~\cite{gross18,beni00,yi11}. The emergence of an equatorial ring of high contractility provides a minimal model for the formation of contractile rings that play a key role during cell division~\cite{whit83,salb12}. Symmetry-breaking instabilities can be biased by external chemical cues. In particular, we found the axis of a contractile ring can be reliably aligned with the axis of a nematic cue, similar to the alignment of a contractile ring with the mitotic spindle axis during cell division~\cite{thery07,rodr15}. 
 
Considering a passive fluid outside of the active surface also leads to the 
formation of patterns via dynamic instabilities. During the polar instability, a net flow outside the surface is generated. Driven by active surface flows, the sphere will therefore move relative to the laboratory frame~\cite{suppPRL}, corresponding to a swimmer that exhibits spontaneous self-propulsion. For the case of vanishing surface viscosity, $\eta_b=\eta_s=0$, this scenario is similar to swimmers driven by Marangoni flows \cite{zoettl16,whit16}. 

Our minimal model captures general features of the contractile actomyosin cortex of cells and its mechanical interactions with the cytoplasm. To account for more complex features of the cell cortex~\cite{salb12}, it could be extended, for example, by including a multi-component descriptions of the biochemical processes. Finally, the instabilities and
surface patterns discussed here could be studied experimentally using \textit{in-vitro}  actomyosin systems reconstituted
in vesicles or in droplets~\cite{murr11,carv13,shah14}.\\

\begin{acknowledgments}
A. Mietke acknowledges funding from an ELBE PhD fellowship. This work was financially supported by the German Federal Ministry of Research and Education, Grant 031L0044 and by the Deutsche Forschungsgemeinschaft (DFG, German Research Foundation) under Germany's Excellence Strategy EXC-2068-390729961, Cluster of Excellence Physics of Life of TU Dresden. V. Jemseena and K. Vijay Kumar acknowledge support from the Max-Planck partner group at the ICTS-TIFR. K. Vijay Kumar is supported by the Department of Biotechnology, India, through a Ramalingaswami reentry fellowship.
\end{acknowledgments}


%

\onecolumngrid
\newpage
\raggedbottom
\setlength{\parindent}{0pt}
\begin{center}
\huge Supplemental Material
\end{center}
\setcounter{page}{1}
\setcounter{equation}{0}
\setcounter{figure}{0}
\renewcommand{\theequation}{S\arabic{equation}}
\renewcommand{\thepage}{S\arabic{page}} 
\renewcommand{\thefigure}{S\arabic{figure}} 
\section{{\large1 Linearization of the hydrodynamic equations}}
The hydrodynamic equations of our deforming active thin film model read
\begin{align}
\nabla_it^{i,d}_{\ j}&=-f^{\text{ext}}_j\label{eq:FB_tang_spec}\\
C^{ij}t_{ij}^d+f_n^e&=f^{\text{ext}}_n,\label{eq:FB_norm_spec}
\end{align} 
where
\begin{align}
t^d_{ij}&=2\eta_s\left(v_{ij}-\frac{1}{2}v^k_{\,k}g_{ij}\right)+\eta_bv^k_{\,k}\,g_{ij}+\xi f(c)g_{ij}\,+\,\xi'g(c)C_{ij}\label{eq:constEq}\\
f_n^e&=2\gamma H+4\kappa\left[2K\left(H-C_0\right)+2H(C_0^2-H^2)-\Delta_{\Gamma}H\right].
\end{align}
Here, $H=C^k_{\,k}/2$ denotes the mean curvature and $f_n^e=0$ corresponds to the shape equation of a Helfrich membrane~\cite{capo02}. In the constitutive Eq.~(\ref{eq:constEq}), we have included a curvature-dependent active tension $\sim\xi'g(c)C_{ij}$ with a stress-regulating function $g(c)$. In general, such a potentially anisotropic contribution  exists in surfaces with up-down broken symmetry~\cite{salb17} and is therefore included in the following linearization. 
External forces $\mathbf{f}^{\text{ext}}=f^{\text{ext}}_i\mathbf{e}^i+f^{\text{ext}}_n\mathbf{n}$ result from the shear stress exerted by the passive Stokes fluid onto the active surface and read
\begin{equation}
\mathbf{f^{\text{ext}}}=\mathbf{n}\left.\cdot\left(\boldsymbol{\bar{\sigma}}-\boldsymbol{\sigma}\right)\right|_{\Gamma}.\label{eq:ExtF}
\end{equation}
Here, $\boldsymbol{\bar{\sigma}}$ and $\boldsymbol{\sigma}$ denote the stress tensor of the passive fluid outside of and enclosed by the surface, respectively, and we included $\boldsymbol{\bar{\sigma}}$ for generality. Viscosities of the passive Stokes fluids inside and outside of the closed surface are denoted $\eta$ and $\bar{\eta}$, respectively.

We linearize the hydrodynamic Eqs.~(\ref{eq:FB_tang_spec}) and (\ref{eq:FB_norm_spec}) around the stationary state in which the active fluid film prescribes the surface of a sphere of radius $R_0$ at rest ($\mathbf{v},\mathbf{u}=0$) and the concentration of stress regulator is homogeneous ($c=c_0$, $\bar{c}=k_{\text{off}}c_0/k_{\text{on}}$). We consider a perturbation expansion in terms of scalar spherical harmonics $Y_{lm}$ and vector spherical harmonics \smash{$\boldsymbol{\Psi}^{(lm)}=R_0\nabla_{\Gamma}Y_{lm}$, $\boldsymbol{\Phi}^{(lm)}=\hat{\mathbf{r}}\times\boldsymbol{\Psi}^{(lm)}$ and $\mathbf{Y}^{(lm)}=Y_{lm}\hat{\mathbf{r}}$} and find from the hydrodynamic equations for each mode $(l,m)$ the system of equations
\begin{align}
&\frac{\eta_s}{R_0^2}(1-l)(l+2)\delta v^{(1)}_{lm}+\frac{\eta_b}{R_0^2}\left(2\delta\dot{R}_{lm}-l(l+1)\delta v^{(1)}_{lm}\right)\nonumber\\
+&\frac{\xi}{R_0}\partial_cf(c_0)\delta c_{lm}+\frac{\xi'}{R_0^2}\left(\partial_cg(c_0)\delta c_{lm}+(l-1)(l+2)g(c_0)\frac{\delta R_{lm}}{R_0}\right)=-\delta f^{(1)}_{lm}\label{eq:LinSpherTang1}\\
&\hspace{7.3cm}\frac{\eta_s}{R_0^2}(1-l)(l+2)\delta v^{(2)}_{lm}=-\delta f^{(2)}_{lm}\label{eq:LinSpherTang2}\\
&\frac{2\eta_b}{R_0^2}\left(2\delta\dot{R}_{lm}-l(l+1)\delta v^{(1)}_{lm}\right)+\frac{2\xi}{R_0}\partial_cf(c_0)\delta c_{lm}+\frac{2\xi'}{R_0^2}\partial_cg(c_0)\delta c_{lm}\nonumber\\
+&\left[\frac{2\kappa}{R_0^4}\left\{l(l+1)-4C_0R_0\right\}+\frac{\gamma+4\kappa C_0^2+\xi f(c_0)}{R_0^2}+\frac{2\xi'g(c_0)}{R_0^3}\right](l-1)(l+2)\delta R_{lm}=\delta f_{lm}^{r},\label{eq:LinSpherNorm}
\end{align}
where dots denote derivatives with respect to time. For each mode, this system determines the coefficients describing in-plane flows ($\delta v_{lm}^{(1)}$, $\delta v_{lm}^{(2)}$) and deformations ($\delta R_{lm}$) as a function of concentration changes ($\delta c_{lm}$). We have furthermore introduced the perturbation-induced changes of the external viscous shear stress as
\begin{equation}
\delta\mathbf{f}^{\text{ext}}=\sum_{l=0}^{\infty}\sum_{m=-l}^{m=l}\left(\delta f_{lm}^{(1)}\boldsymbol{\Psi}^{(lm)}+\delta f_{lm}^{(2)}\boldsymbol{\Phi}^{(lm)}+\delta f_{lm}^{r}\mathbf{Y}^{(lm)}\right),\label{eq:ExtF2}
\end{equation}
where
\begin{align}
R_0\delta f_{lm}^{(1)}=&\,-(\eta+\bar{\eta})(1+2l)\delta v_{lm}^{(1)}+3[\eta(l+1)+\bar{\eta}l]\frac{\delta\dot{R}_{lm}}{l(l+1)}+\frac{\bar{\eta}}{2}A\delta_{l,1}\label{eq:stressPsi}\\
R_0\delta f_{lm}^{(2)}=&\,-\left[\eta(l-1)+\bar{\eta}(l+2)\right]\delta v_{lm}^{(2)}\label{eq:stressPhi}\\
R_0\delta f_{lm}^{r}=&\,3\left[\eta(l+1)+\bar{\eta}l\right]\delta v_{lm}^{(1)}-\left[\left(\eta+\bar{\eta}\right)\left(4+3l+2l^2\right)l+3\eta\right]\frac{\delta\dot{R}_{lm}}{l(l+1)}+\frac{\bar{\eta}}{2}A\delta_{l,1}\label{eq:stressY},
\end{align}
and $A=2\delta v_{lm}^{(1)}+\delta\dot{R}_{lm}$. Equations~(\ref{eq:stressPsi})--(\ref{eq:stressY}) follow from an analytic solution of the Stokes equation~\cite{seyb18} for no-slip and impermeability boundary conditions at the surface [Eqs.~(9)--(10)]. 

The linearization of the convection-diffusion Eq.~(11) reads
\begin{align}
\delta\dot{c}_{lm}+\frac{c_0}{R_0}\left(2\delta\dot{R}_{lm}-l(l+1)\delta v^{(1)}_{lm}\right)+\left(\frac{D}{R_0^2}l(l+1)+k_{\text{off}}\right)\delta c_{lm}&=0.\label{eq:LinSpherCont}
\end{align}

\section*{{\large2 Determining the critical P\'eclet number}}
In the following, we describe how a critical P\'eclet number can be derived from the linearized Eqs.~(\ref{eq:LinSpherTang1})--(\ref{eq:LinSpherNorm}). While we introduced a more general system in Sec.~1, the critical P\'eclet number $\text{Pe}^*_l$ given in Eq.~(13) (main text) corresponds to the case, where the passive fluid viscosity outside of the closed surface and the curvature-dependent active tension vanish, i.e. $\bar{\eta}=0$ and $\xi'=0$.  

\subsection*{2.1 General procedure and treatment of the translational mode}
First, we use Eqs.~(\ref{eq:LinSpherTang1})--(\ref{eq:LinSpherNorm}) to eliminate the mode coefficients for the in-plane flows: From Eq.~(\ref{eq:LinSpherTang2}) we find $\delta v^{(2)}_{lm}=0$ for $l\ge2$. Additionally, we exclude full body rotations, such that $\delta v^{(2)}_{lm}=0$ for all $l$. The remaining equations can be used to eliminate $\delta v^{(1)}_{lm}$ from Eqs.~(\ref{eq:LinSpherNorm}) and (\ref{eq:LinSpherCont}) to arrive at expressions for $\delta\dot{R}_{lm}$ and $\delta\dot{c}_{lm}$ in terms of $\delta c_{lm}$ and $\delta R_{lm}$, effectively yielding the Jacobian $\mathcal{J}_l$ associated with this perturbation. Note that for $l=1$,  Eqs.~(\ref{eq:LinSpherTang1}) and (\ref{eq:LinSpherNorm}) determine only the net in-plane compression $\nabla_{\Gamma}\cdot\delta\mathbf{v}|_{l=1}\sim\delta\dot{R}_{1m}-\delta v^{(1)}_{1m}$ due to the presence of translational modes that we have not fixed so far. Any such fix has to imply a relation between $\delta v^{(1)}_{1m}$ and $\delta\dot{R}_{1m}$. However, because the linearized convection-diffusion Eq.~(\ref{eq:LinSpherCont}) also depends for $l=1$ exclusively on the compression rate $\sim\delta\dot{R}_{1m}-\delta v^{(1)}_{1m}$, the dispersion relation, and consequently the critical P\'eclet number, is independent of the constraint by which translational modes are fixed. 

\subsection*{2.2 Instabilities of the polar mode $l=1$}
Shape perturbations of the mode $l=1$ to first order do not change the curvature of a spherical surface. As a consequence, the linearized Eqs.~(\ref{eq:LinSpherTang1})--(\ref{eq:LinSpherNorm}) drastically simplify. Furthermore, the curvature $C_{ij}$ is to zeroth order isotropic, such that contributions from $\xi$ and $\xi'$ are essentially equivalent for $l=1$, if we identify $\xi f(c)\leftrightarrow \xi'g(c)/R_0$.

Instabilities of the polar mode $l=1$ occur for increasing P\'eclet number at 
\begin{equation}
\text{Pe}^*_{l=1}=\frac{1}{c_0\partial_cf(c_0)}\left(1+\frac{1}{2}\tau_Dk_{\text{off}}\right)\left[2+\nu\left(\frac{3R_0}{L_h}+\frac{2R_0}{\bar{L}_h}\right)\right].\label{eq:Pesupp}
\end{equation}
In addition to the hydrodynamic length $L_h=\eta_s/\eta$, we have introduced here $\bar{L}_h=\eta_s/\bar{\eta}$ as the hydrodynamic screening length associated with the passive fluid outside of the closed surface. In the limit $\bar{L}_h\rightarrow\infty$, when the viscosity of the passive fluid outside of the closed surface can be neglected, Eq.~(\ref{eq:Pesupp}) is equivalent to Eq.~(13) with $l=1$ (main text).

\subsection*{2.3 Components of the Jacobian for $l\ge2$}
Following the general procedure outlined in Sec.~2.1, we derive in the following the components of the system's Jacobian. The Jacobian diagonalizes in the space of spherical harmonics and contributes for each mode $l$ with
\begin{equation}
\mathcal{J}_l=\begin{pmatrix}\mathcal{J}^l_{RR} & \mathcal{J}^l_{Rc}\\\mathcal{J}^l_{cR} & \mathcal{J}^l_{cc}\end{pmatrix}.\label{eq:GenJac}
\end{equation}
To determine its components, we first use Eqs.~(\ref{eq:LinSpherTang1}) and (\ref{eq:LinSpherNorm}) and find:
\begin{align}
\delta v_{lm}^{(1)}=\Lambda_1\delta R_{lm}+\Lambda_2\delta\dot{R}_{lm},\label{eq:dvgen}
\end{align}
where
\begin{align}
\Lambda_1&=\frac{(l+2)(1-l)\left[2\kappa l(l+1)/R_0^2+\hat{\gamma}+\xi f(c_0)\right]}{(l-1)R_0\eta+(l+2)R_0\bar{\eta}+2(l-1)(l+2)\eta_s}\label{eq:L1}\\
\Lambda_2&=\frac{3\eta-l(l+2)(2l-1)(\eta+\bar{\eta})}{l(l+1)\left[(l-1)\eta+(l+2)\bar{\eta}+2(l-1)(l+2)\eta_s/R_0\right]}.\label{eq:L2}
\end{align}
In Eqs.~(\ref{eq:L1}) and (\ref{eq:L2}), we have collected contributions to an effective passive surface tension into
\begin{equation}
\hat{\gamma}=\gamma+4\kappa C_0(C_0-2/R_0).\label{eq:passST}
\end{equation}
Next, we use Eqs.~(\ref{eq:LinSpherTang1}) and Eq.~(\ref{eq:dvgen}), which can be rearranged into
\begin{equation}
\delta\dot{R}_{lm}=\frac{1}{\Gamma}\left[\left(\Lambda_1J_{RR}+J_{RR}'\right)\delta R_{lm}+\left(J_{Rc}+J_{Rc}'\right)\delta c_{lm}\right]\label{eq:dRgen}
\end{equation}
with
\begin{align}
\Gamma&=\frac{\eta_s}{R_0}(l-1)(l+2)\Lambda_2+\frac{\eta_b}{R_0}\left[l(l+1)\Lambda_2-2\right]\nonumber\\
&+\left(\eta+\bar{\eta}\right)(1+2l)\Lambda_2-3\left(\frac{\eta}{l}+\frac{\bar{\eta}}{l+1}\right)\\
J_{RR}&=-\frac{\eta_s}{R_0}(l-1)(l+2)-\frac{\eta_b}{R_0}l(l+1)-\left(\eta+\bar{\eta}\right)(1+2l)\label{eq:JRR}\\
J_{Rc}&=\xi\partial_cf(c_0)\label{eq:JRC}\\
J'_{RR}&=\frac{\xi'g(c_0)}{R_0^2}(l-1)(l+2)\\
J'_{Rc}&=\frac{\xi'}{R_0}\partial_cg(c_0).
\end{align}
From Eq.~(\ref{eq:dRgen}), one can read off the components of the Jacobian that are defined by $\dot{R}_{lm}=\mathcal{J}^l_{RR}\delta R_{lm}+\mathcal{J}^l_{Rc}\delta c_{lm}$. Finally, we use Eqs.~(\ref{eq:LinSpherCont}), (\ref{eq:dvgen}) and (\ref{eq:dRgen}) to write
\begin{equation}
\delta\dot{c}_{lm}=\left(J_{cc}+J_{cc}'\right)\delta c_{lm}+\left(\Lambda_1J_{cR}+J_{cR}'\right)\delta R_{lm},\label{eq:dcgen}
\end{equation}
where
\begin{align}
J_{cc}&=-\left(\frac{D}{R_0^2}l(l+1)+k_{\text{off}}+\frac{\xi c_0\partial_cf}{\Gamma R_0}\left[2-l(l+1)\Lambda_2\right]\right)\\
J_{cR}&=\frac{c_0}{R_0}\left[l(l+1)\left(1+\frac{\Lambda_2}{\Gamma}J_{RR}\right)-\frac{2}{\Gamma}J_{RR}\right]\\
J'_{cc}&=-\frac{\xi'c_0\partial_cg}{\Gamma R_0^2}\left[2-l(l+1)\Lambda_2\right]
\end{align}
\begin{align}
J'_{cR}&=-\frac{\xi'g(c_0)c_0}{\Gamma R_0^3}\left[2-l(l+1)\Lambda_2\right](l-1)(l+2).\label{eq:Jcrp}
\end{align}
From Eq.~(\ref{eq:dcgen}), one can read off the remaining components of the Jacobian defined by $\delta\dot{c}_{lm}=\mathcal{J}^l_{cR}\delta R_{lm}+\mathcal{J}^l_{cc}\delta c_{lm}$.\\

\subsection*{2.4 Critical parameter values for $l\ge2$}
From the Jacobian $\mathcal{J}_l$ given in Eq.~(\ref{eq:GenJac}) and its components defined in Eqs.~(\ref{eq:dvgen})--(\ref{eq:Jcrp}), we now determine for a given mode $l$ the critical parameter values for which the system becomes unstable. A numerical analysis of the Jacobian and the comparison with its asymptotic behavior for large bending rigidities indicates that instability transitions are typically stationary. Critical parameter values can therefore be determined from the condition det$\left(\mathcal{J}_l\right)=0$. The computation of the determinant yields
\begin{align}
-\Gamma\det\left(\mathcal{J}_l\right)&=\left(\Lambda_1J_{RR}+\frac{\xi'g(c_0)}{R_0^2}(l-1)(l+2)\right)\left(\frac{D}{R_0^2}l(l+1)+k_{\text{off}}\right)\nonumber\\
&+\Lambda_1\frac{\left[R_0\xi\partial_cf(c_0)+\xi'\partial_cg(c_0)\right]c_0}{R_0^2}l(l+1).
\end{align}

In addition to the P\'eclet number defined in the main text, Pe\,$=\xi R_0^2/(D\eta_b)$, we define an analogue dimensionless quantity associated with the curvature-dependent active tension: Pe$'=\xi'R_0/(D\eta_b)$. The mode $l$ is unstable if the P\'eclet numbers Pe and Pe$'$ fulfill the condition 
\begin{align}
&\text{Pe}c_0\partial_cf(c_0)+\text{Pe}'c_0\partial_cg(c_0)>\nonumber\\
&\left(1+\frac{\tau_Dk_{\text{off}}}{l(l+1)}\right)\left[l(l+1)+\nu\left((l-1)(l+2)+(1+2l)\frac{R_0}{L^{\text{eff}}_h}+\frac{\text{Pe}'g(c_0)}{\tilde{\Lambda}_1}\right)\right].\label{eq:StabCond}
\end{align}
Here, $\tau_D=R_0^2/D$, $L_h^{\text{eff}}=L_h\bar{L}_h/(L_h+\bar{L}_h)$ is an effective hydrodynamic length, and $\tilde{\Lambda}_1$ is derived from Eq.~(\ref{eq:L1}) as
\begin{equation}
\tilde{\Lambda}_1=\frac{\frac{\tau_D}{\eta_b}\left[2\kappa l(l+1)/R_0^2+\hat{\gamma}\right]+\text{Pe}f(c_0)}{(l-1)R_0/L_h+(l+2)R_0/\bar{L}_h+2(l-1)(l+2)},
\end{equation}
where $\hat{\gamma}$ is defined in Eq.~(\ref{eq:passST}). \\

From Eq.~(\ref{eq:StabCond}) it follows that larger gradients of the stress regulator functions,  $\partial_cf(c_0)$ and $\partial_cg(c_0)$, both tend to destabilize the homogeneous state for given $\text{Pe}>0$ and $\text{Pe}'>0$ [left-hand side of Eq.~(\ref{eq:StabCond})]. Furthermore, the steady-state values of the active tension $\sim\text{Pe}f(c_0)$ and $\sim\text{Pe}'g(c_0)$ affect the stability of the homogeneous state, which is not the case if $\text{Pe}'=0$. In particular, in conjunction with the curvature-dependent active tension, an increase in active steady-state tension $\sim\text{Pe}f(c_0)$ tends to destabilize the homogeneous state [right-hand side of Eq.~(\ref{eq:StabCond})].

For larger bending rigidity or surface tension, $\kappa\gg R_0^2\eta_b/\tau_D$ or $\hat{\gamma}\gg\eta_b/\tau_D$, we have $\tilde{\Lambda}_1\gg1$ and the last term in Eq.~(\ref{eq:StabCond}) can be neglected, which leads to the prediction of instabilities for
\begin{align}
&\text{Pe}c_0\partial_cf(c_0)+\text{Pe}'c_0\partial_cg(c_0)>\nonumber\\
&\left(1+\frac{\tau_Dk_{\text{off}}}{l(l+1)}\right)\left[l(l+1)+\nu\left((l-1)(l+2)+(1+2l)\frac{R_0}{L^{\text{eff}}_h}\right)\right].\label{eq:StabCondLim}
\end{align}

In this regime, contributions from the isotropic active tension and from the curvature-dependent active tension yield the same physics in terms of the linear instabilities. Furthermore, Eq.~(\ref{eq:StabCondLim}) represents the instability criterion in the limit where the surface is a rigid sphere. This can be checked independently by directly deriving the dispersion relation for the dynamics on a rigid sphere using Eqs.~(\ref{eq:LinSpherTang1}) and (\ref{eq:LinSpherCont}) for  $\delta R_{lm}=0$ and $\delta \dot{R}_{lm}=0$.

\subsection*{2.5 Critical P\'eclet number and eigenmode components for isotropic active tension}
In the absence of a curvature-dependent active tension [Pe$'=0$ in Eq.~(\ref{eq:StabCond})] the critical P\'eclet number Pe$^*_l$ reads for $l\ge2$:
\begin{equation}
\text{Pe}^*_{l\ge2}=\frac{1}{c_0\partial_cf(c_0)}\left(1+\frac{\tau_Dk_{\text{off}}}{l(l+1)}\right)\left[l(l+1)+\nu\left((l-1)(l+2)+(1+2l)\frac{R_0}{L^{\text{eff}}_h}\right)\right].\label{eq:Pel}
\end{equation}
This expression is equivalent to Eq.~(13) in the main text in the limit $\bar{L}_h\rightarrow\infty$, when the viscosity of passive fluid outside of the closed surface vanishes.

From Eq.~(\ref{eq:Pel}) it follows that instabilities of modes with $l\ge2$ can be characterized by an effective hydrodynamic length $L_h^{\text{eff}}=L_h\bar{L}_h/(L_h+\bar{L}_h)$ or, alternatively, by an effective viscosity $\eta^{\text{eff}}=\eta+\bar{\eta}$ of the surrounding passive fluids. Mechanical interactions with a passive fluid outside of the active surface therefore give rise to the same phenomenology as for the enclosed passive fluid case discussed in the main text. This is because both cases lead to hydrodynamic screening length on the surface, which is one of the key requirements for the described patterning mechanism. Furthermore, in the absence of a curvature-dependent active tension, mechano-chemical instability transitions are generally independent of surface tension~$\gamma$, bending rigidity~$\kappa$ and spontaneous curvature $C_0$.

Note that in the limit of large surface tension or bending rigidity the surface remains a non-deforming sphere. In this case, the condition of a force-free surface dynamics with $\bar{\eta}\ne0$, i.e. $\oint_{\Gamma}dA\,\boldsymbol{\bar{\sigma}}\cdot\mathbf{n}=0$, implies for any surface flow $\sim\delta v_{1m}$ a net motion of the surface relative to the laboratory frame~\cite{zoettl16} (see also Sec.~4.3). Hence, a polar instability at the critical P\'eclet number given in Eq.~(\ref{eq:Pesupp}) is for $\bar{\eta}\ne0$ accompanied by the spontaneous onset of well-defined translational motion.\\

Finally, we compute the components of the critical eigenmodes $\boldsymbol{\delta}_{lm}^*$ for $l\ge2$. Using the fact that the critical P\'eclet number follows from det$\left(\mathcal{J}_l\right)=0$, the critical eigenmode must be of the form
\begin{align}
\begin{pmatrix}
\delta R_{lm}^*/R_0\\
\delta c_{lm}^*/c_0
\end{pmatrix}
=
\frac{1}{N}
\begin{pmatrix}
\mathcal{J}^l_{Rc}/(R_0\mathcal{J}^l_{RR})\\
-1
\end{pmatrix},\label{eq:eigenmvec}
\end{align}
where $N$ denotes some normalizing prefactor. The eigenmode component related to deformations is therefore proportional to
\begin{align}
\frac{\mathcal{J}^l_{Rc}}{R_0\mathcal{J}^l_{RR}}&=\frac{(l-1)R_0\eta+(l+2)R_0\bar{\eta}+2(l-1)(l+2)\eta_s}{(l+2)(l-1)\left[2\kappa l(l+1)/R_0^2+\hat{\gamma}+\xi f(c_0)\right]}\nonumber\\
&\times\frac{\xi c_0\partial_cf(c_0)}{\left[(l-1)(l+2)\eta_s+l(l+1)\eta_b+(1+2l)R_0\left(\eta+\bar{\eta}\right)\right]},\label{eq:eigenm}
\end{align}
which follows from Eqs.~(\ref{eq:GenJac}), and (\ref{eq:dRgen})--(\ref{eq:JRC}). Equation~(\ref{eq:eigenm}) has two important implications: First, in the limit of large bending rigidity (or surface tension), we have $\delta R_{lm}^*\sim1/\kappa\rightarrow0$ (see Fig.~1\,c). Second, we note that the expression on the right-hand side of Eq.~(\ref{eq:eigenm}) is manifestly positive. Together with the form of the eigenmode Eq.~(\ref{eq:eigenmvec}), this provides a formal proof that an increase of stress regulator around the equator ($\delta c^*_{2,0}<0$) induces a shape constriction around the equator ($\delta R_{2,0}^*>0$) if the mode $l=2, m=0$ becomes unstable and a contractile ring forms (see Fig.~1\,b inset).

\section{{\large3 Nonlinear dynamics on the surface of a rigid sphere}}
\subsection*{3.1 Pattern formation and dynamics beyond the linear instability}
\begin{figure}[t]
	\centering	
\includegraphics[width=0.5\textwidth]{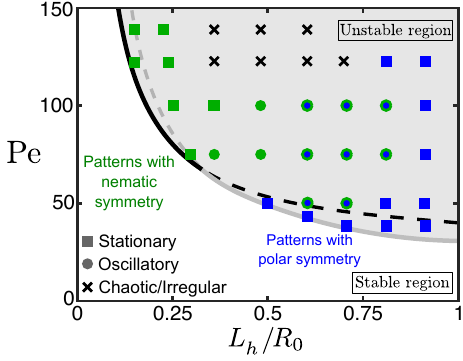} 
\caption{Pattern formation and dynamics in the unstable region of the stability diagram (same parameters as in Fig.~1\,b, main text) in the limit of large bending rigidity (rigid sphere). Beyond the linear instability threshold stationary patterns (squares) with the predicted symmetry emerge (patterns shown in Fig.~2, main text). Further away from the instability threshold oscillatory patterns can form (disks). Blue/Green disks: a single patch of high stress regulator concentration propagates in a fixed equatorial plane around the surface, corresponding to a mixed polar and nematic symmetry. Green disks: A nematic pattern oscillates between a centered contractile ring and a pair of high concentration patches at opposite poles. Black crosses: No distinct symmetry can be identified due to an irregular pattern dynamics. \label{fig:S1}}
\end{figure}

Using numerical simulations, we analyzed the pattern formation in the limit of large bending rigidity (rigid sphere) across the whole unstable region of the stability diagram shown in Fig.~1\,b (main text). Surface patterns of the expected polar and nematic symmetry form robustly near the instability thresholds~(squares in Fig.~\ref{fig:S1}). Further away, numerical simulations additionally reveal the existence of oscillatory steady state patterns~(disks in Fig.~\ref{fig:S1}). 
\subsection*{3.2 Instabilities of higher harmonic modes}
We have focused in our work on the formation of polar and nematic patterns as the most relevant cases in the context of biological processes. However, the general hydrodynamic screening also allows for the emergent formation of surface patterns with symmetries corresponding to higher harmonics modes. Such patterns will form if a higher harmonic mode becomes the only unstable mode, which occurs at even smaller hydrodynamic lengths and larger P\'eclet numbers than discussed in the main text. Depending on the parity of the unstable harmonic mode, the resulting steady state patterns of higher modes also have polar (odd modes) or nematic symmetry (even modes). Two examples of stationary states corresponding to instabilities of the modes $l=3$ and $l=4$ are shown in Fig.~S2.
\begin{figure}[t]
	\centering	
\includegraphics[width=0.65\textwidth]{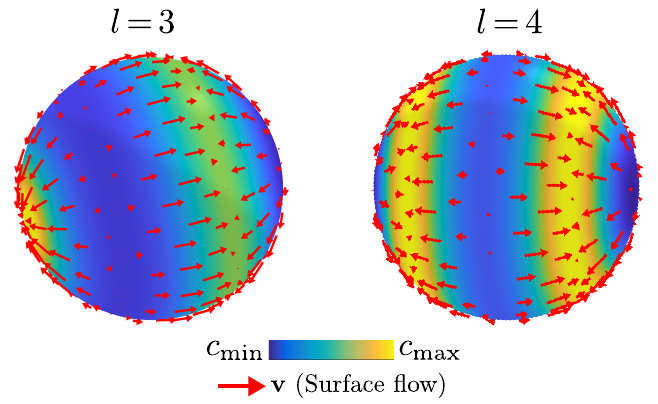} 
\caption{ Stationary patterns in a regime where the mode $l=3$ (left) or $l=4$ (right) are the only unstable modes. Parameters: $\text{Pe}=315$, $L_h/R_0=1/15.5$, $k\tau_D=16$ ($l=3$), $\text{Pe}=610$, $L_h/R_0=1/24$, $k\tau_D=27$ ($l=4$) and $\nu=1$. Active tension is regulated by $f(c)=2c^2/(c_0^2+c^2)$, such that $c_0\partial_cf(c_0)=1$. \label{fig:S2}}
\end{figure}

\subsection*{3.3 General analytic solution for active surface flows on a fixed sphere}
The in-plane hydrodynamic Eq.~(\ref{eq:FB_tang_spec}) on a fixed sphere is linear in the surface flows $\mathbf{v}_{\parallel}=v^i\mathbf{e}_i$ and a closed analytic solution can be derived. This solution forms the basis of the numerical approach that we have used to solve the fully non-linear problem on a fixed sphere (Sec.~3.4). Note that on a rigid sphere the curvature is isotropic, such that active contributions $\sim\xi$ and $\sim\xi'$ in Eq.~(\ref{eq:constEq}) are equivalent and we include w.l.o.g. only~$\xi$ in the following derivation. To derive the analytic solution, we first note that on an arbitrary non-deforming surface the hydrodynamic Eq.~(6) (main text) can be expressed as
\begin{equation}\label{eq:FB_iso_rs}
\eta_s\left(\nabla_i\nabla^iv_j+Kv_j\right)+\eta_b\nabla_j\nabla_iv^i+\xi\partial_jf(c)=-f^{\text{ext}}_j,
\end{equation}
where $K=\det\left(C_i^{,j}\right)$ is the Gaussian curvature of the surface and we have used the Ricci identity on two-dimensional surfaces~\cite{sand78}: $\nabla_i\nabla_jv^i-\nabla_j\nabla_iv^i=Kv_j$. Expanding surface flows on the fixed sphere as
\begin{align}
\mathbf{v}_{\parallel}&=\sum_{l=0}^{\infty}\sum_{m=-l}^{m=l}v^{(1)}_{lm}\boldsymbol{\Psi}^{(lm)}\label{eq:vexp_supp},
\end{align}
and the distribution of active tension as
\begin{align}
f(c)&=\sum_{l=0}^{\infty}\sum_{m=-l}^{m=l}f_{lm}Y_{lm}\label{eq:fofc},
\end{align}
Eq.~(\ref{eq:FB_iso_rs}) implies
\begin{align}\label{eq:v_iso}
v_{lm}^{(1)}&=\frac{R_0\xi}{R_0\left[\left(\eta+\bar{\eta}\right)(1+2l)-\bar{\eta}\delta_{l,1}\right]+l(l+1)(\eta_b+\eta_s) -2\eta_s}f_{lm}.
\end{align}
Here, we have used covariantly formulated properties of vector spherical harmonics that read on the unit sphere~\cite{sand78}: $\Psi^{(lm)}_i=\partial_iY_{lm}$,  $\nabla^i\Psi^{(lm)}_i=-l(l+1)Y_{lm}$ and $\nabla_i\nabla^i\Psi^{(lm)}_j=\left[1-l(l+1)\right]\Psi^{(lm)}_j$. Furthermore, we have used that $f^{\text{ext}}_j$ is given by Eqs.~(\ref{eq:ExtF2})--(\ref{eq:stressPhi}), which become exact expressions on a fixed sphere. Equation~(\ref{eq:v_iso}) provides an analytic solution for surfaces flows $\mathbf{v}_{\parallel}$ that result from an arbitrary distribution of active isotropic tension $\sim\xi f(c)g_{ij}$ on the surface. Note that the solution Eq.~(\ref{eq:v_iso}) can also be read off from Eq.~(\ref{eq:LinSpherTang1}) for $\delta\dot{R}_{lm}=0$, $\delta v_{lm}^{(1)}\rightarrow v_{lm}^{(1)}$ and $\partial_cf\delta c_{lm}\rightarrow f_{lm}$. 

The flow and pressure of the passive fluid driven by active surface flows are given by 
\begin{align}
\mathbf{u}&=\sum_{l=0}^{\infty}\sum_{m=-l}^{m=l}\left(u_{lm}^{(1)}(r)\boldsymbol{\Psi}^{(lm)}+u_{lm}^{r}(r)\mathbf{Y}^{(lm)}\right)\\
p&=\sum_{l=0}^{\infty}\sum_{m=-l}^{m=l}p_{lm}(r)Y_{lm},
\end{align}
where the coefficient functions are given by~\cite{seyb18}\ \\
\begin{align}
u_{lm}^{(1)}(r)=&\,A_{lm}^{(1)}\frac{(l+3)}{l(l+1)}\left(\frac{r}{R_0}\right)^{l+1}+\frac{A_{lm}^{(2)}}{l}\left(\frac{r}{R_0}\right)^{l-1}+\bar{A}_{1m}^{(2)}\delta_{l,1}\nonumber\\
-&\,A_{lm}^{(3)}\frac{l-2}{l(l+1)}\left(\frac{r}{R_0}\right)^{-l}-\frac{A_{lm}^{(4)}}{l+1}\left(\frac{r}{R_0}\right)^{-l-2}\label{eq:u1_lm}
\end{align}
\begin{align}
u_{lm}^r(r)=&\,A_{lm}^{(1)}\left(\frac{r}{R_0}\right)^{l+1}+A_{lm}^{(2)}\left(\frac{r}{R_0}\right)^{l-1}+\bar{A}_{1m}^{(2)}\delta_{l,1}+A_{lm}^{(3)}\left(\frac{r}{R_0}\right)^{-l}+A_{lm}^{(4)}\left(\frac{r}{R_0}\right)^{-l-2}\label{eq:ur_lm}\\
p_{lm}(r)=&\,\eta A_{lm}^{(1)}\frac{4l+6}{l}\frac{r^l}{R_0^{l+1}}+\bar{\eta}A_{lm}^{(3)}\frac{4l-2}{l+1}\frac{r^{-l-1}}{R_0^{l}}.
\end{align}
Here, the coefficients $A_{lm}^{(1)}$ and $A_{lm}^{(2)}$ yield the solution of the Stokes equation inside the sphere, $A_{lm}^{(3)}$ and $A_{lm}^{(4)}$ outside the sphere. For $l=1$, the outside solution contains in general a third coefficient $\bar{A}_{1m}^{(2)}$ related to translational motion. The latter is fixed by imposing a force-free surface dynamics, $\oint_{\Gamma}dA\,\boldsymbol{\bar{\sigma}}\cdot\mathbf{n}=0$, which is equivalent to $A_{1m}^{(3)}=0$. For impermeability and no-slip boundary conditions, a solution \textit{in the rest-frame} of the non-deforming sphere is then given by $u_{lm}^r(R_0)=0$ and $u_{lm}^{(1)}(R_0)=v_{lm}^{(1)}$, which implies
\begin{align}
A_{lm}^{(1)}&=\frac{1}{2}l(1+l)v_{lm}^{(1)}\\
A_{lm}^{(2)}&=-\frac{1}{2}l(1+l)v_{lm}^{(1)}\\
A_{lm}^{(3)}&=\frac{1}{2}l(1+l)v_{lm}^{(1)}-\delta_{l,1}v_{lm}^{(1)}\label{eq:A3}\\
A_{lm}^{(4)}&=\frac{1}{2}l(1+l)v_{lm}^{(1)}+\frac{1}{3}\delta_{l,1}v_{lm}^{(1)}.\label{eq:A4}
\end{align}
and $\bar{A}^{(2)}_{1m}=2v_{1m}^{(1)}/3\sim|\mathbf{u}(r\rightarrow\infty)|$. Because this solution was computed in the rest-frame of the sphere, the constant flow field at infinity described by $\bar{A}^{(2)}_{1m}$ corresponds to a translational motion relative to the laboratory frame~\cite{zoettl16}.

Writing the concentration field of the stress regulator as
\begin{align}\label{eq:cExp}
c=\sum_{l=0}^{\infty}\sum_{m=-l}^{m=l}c_{lm}Y_{lm},
\end{align}
the advection-diffusion Eq.~(11) (main text) implies on a fixed sphere for each mode 
\begin{equation}\label{eq:cDyn}
\frac{d}{dt}c_{lm}=-\frac{D}{R_0^2}l(l+1)c_{lm}+J_{lm}+a_{lm}.
\end{equation}
Here, $J_{lm}$ denotes the harmonic expansion of the exchange of molecules between the thin film and the enclosed fluid [Eqs.~(12) and (14)] given by
\begin{equation}
J_{lm}=\left\{
\begin{array}{l}
2\sqrt{\pi}k_{\text{on}}\bar{c}\,\delta_{l,0}-k_{\text{off}}c_{lm}\text{\hspace{2.5cm (\textit{without} nematic cue)}}\\
2\sqrt{\pi}k^{(0)}_{\text{on}}\bar{c}\left(\delta_{l,0}-\beta\frac{2}{5}\delta_{l,2}\right)-k_{\text{off}}c_{lm} \text{\hspace{0.5cm (\textit{with} nematic cue)}}
\end{array}
\right.
\end{equation}
Throughout, we have parametrized the homogeneous bulk concentration as $\bar{c}=k_{\text{off}}c_0/k_{\text{on}}$ $(=k_{\text{off}}c_0/k^{(0)}_{\text{on}})$ and used $c_0$ as the characteristic surface concentration of the system. The coefficients $a_{lm}$ in Eq.~(\ref{eq:cDyn}) result from the nonlinear advection term in the advection-diffusion equation, and they are given by
\begin{align}
a_{lm}=&-\int\nabla_i(cv^i)Y_{lm}^*d\Omega\nonumber\\
&=\frac{1}{2R_0}\sum_{l_1,l_2,m_1,m_2}c_{l_1m_1}^{\,}v^{(1)}_{l_2m_2}\left[l(l+1)+l_2(l_2+1)-l_1(l_1+1)\right]\int Y_{l_1m_1}Y_{l_2m_2}Y_{lm}^*d\Omega.\label{eq:Gaunt}
\end{align}
Here, $d\Omega=\sin\theta d\theta d\varphi$ denotes the spherical solid angle and the integrals over products of three spherical harmonics are known as Gaunt coefficients~\cite{gaunt29}. 

\subsection*{3.4 Numerical solution}
Based on the analytic solution introduced above, we have implemented a spectral solver that uses a least-squared formulation to approximate all spatial dependencies in terms of spherical harmonics~\cite{snee94}. These projections are required to determine the coefficients $f_{lm}$ in Eq.~(\ref{eq:fofc}) for a given concentration field $c$, as well as the coefficients $a_{lm}$ [Eq.~(\ref{eq:Gaunt})]. The latter can be found for given \smash{$v^{(1)}_{lm}$} and $c_{lm}$ either explicitly from Eq.~(\ref{eq:Gaunt}) or by first calculating numerically in real space the scalar field $a=R_0\nabla_{i}\left(cv^i\right)$ from
\begin{align}\label{eq:AdvProd}
a&=R_0\left(c\nabla_iv^i+\mathbf{v}_{\parallel}\cdot\nabla_{\Gamma}c\right)\nonumber\\
&=-c\sum_{l=0}^{\infty}\sum_{m=-l}^{m=l}l(l+1)v^{(1)}_{lm}Y_{lm}+\sum_{l,m}c_{lm}\mathbf{v}_{\parallel}\cdot\boldsymbol{\Psi}^{(lm)}
\end{align}
using $c$ and $\mathbf{v}_{\parallel}$ given in Eqs.~(\ref{eq:vexp_supp}) and (\ref{eq:cExp}), respectively, and second determining the coefficients $a_{lm}$ as the numerical least-squared harmonic projection of $a$.

To obtain numerical solutions in practice, we consider a concentration field as given in Eq.~(\ref{eq:cExp}), where we take all harmonic modes up to $l=16$, i.e. 289 modes overall, into account. We then determine $f(c)$ and its harmonic expansion coefficients $f_{lm}$ using the harmonic least-squared projection~\cite{snee94}. The harmonic mode coefficients corresponding to the surface and bulk flows are analytically given by Eq.~(\ref{eq:v_iso}), and by Eqs.~(\ref{eq:u1_lm}) and (\ref{eq:ur_lm}), respectively. With this, all terms on the right-hand side of Eq.~(\ref{eq:cDyn}) can be determined from given concentration modes $c_{lm}$. The time integration of Eq.~(\ref{eq:cDyn}) is performed using the Dormand-Prince method (RKDP)~\cite{dorm80} as implemented in Matlab~\cite{matl}.  

\end{document}